\documentclass[runningheads]{llncs}
\usepackage[T1]{fontenc}
\usepackage{graphicx}
\usepackage[
  colorlinks=true, citecolor=red, pdfborder={0 0 0}
]{hyperref}
\newcommand{\cmmnt}[1]{\ignorespaces}

\newcommand{\ti}[1]{\textit{#1}}

\newcommand{\fig}[3][0.95]{
    \begin{figure}[tb]
        \centering\includegraphics[width=#1\linewidth]{sub-images/#2}
        \vspace{-10pt}\caption{#3}
        \vspace{-10pt}\label{fig:#2}
    \end{figure}
}
\usepackage{color}

\begin{document}
\title{Exploring Urban Mobility Trends Using Cellular Network Data
}
%
%
\author{Oluwaleke Yusuf\inst{1}\orcidID{0000-0002-5904-648X} \and
    Adil Rasheed\inst{1}\orcidID{0000-0003-2690-983X} \and
    Frank Lindseth\inst{2}\orcidID{0000-0002-4979-9218}
}
\authorrunning{O. Yusuf et al.}
%
\institute{Department of Engineering Cybernetics \\
    \email{\{oluwaleke.u.yusuf, adil.rasheed\}@ntnu.no}
    \and Department of Computer Science \\
    \email{frankl@ntnu.no} \\
    Norwegian University of Science and Technology, 7034 Trondheim, Norway
}
\maketitle

\begin{abstract}
    The growth of urban areas intensifies the need for sustainable, efficient transportation infrastructure and mobility systems, driving initiatives to enhance infrastructure and public transit while reducing traffic congestion and emissions. By utilizing real-world data, a data-driven approach can provide crucial insights for urban mobility planning and decision-making.
    This study explores the efficacy of leveraging telecoms data from cellular network signals for studying crowd movement patterns, focusing on Trondheim, Norway. It examines routing reports to understand the spatiotemporal dynamics of various transportation routes and modes.
    A data preprocessing and feature engineering framework was developed to process raw routing reports for historical analysis. This enabled the examination of geospatial trends and temporal patterns, including a comparative analysis of various transportation modes, along with public transit usage. Specific routes and areas were analyzed in-depth to compare their mobility patterns with the broader city context.
    The study highlights the potential of cellular network data as a resource for shaping urban transportation and mobility systems. By identifying deficiencies and potential improvements, city planners and stakeholders can foster more sustainable and effective transportation and mobility solutions.
\end{abstract}

\section{Introduction}
The expansion of urban centers such as Trondheim, Norway's third-largest city with a 2023 population of approximately 206,000 \cite{SSB04861}, underscores the pressing need for infrastructure modernization to address challenges such as traffic congestion and pollution driven by rising populations and mobility demands. Trondheim's growth---spurred by the presence of educational and research institutions like the Norwegian University of Science and Technology, SINTEF, and St. Olav's University Hospital---places pressure on its transportation infrastructure and mobility system. This development has led to a shift toward sustainable mobility solutions, with initiatives like \href{https://miljopakken.no/}{Miljøpakken} and \href{https://mobilitetslabstortrondheim.no/}{MobilitetsLab Stor-Trondheim} (MoST) emphasizing public transit and active transportation over private car usage. Achieving such goals necessitates a data-driven approach toward understanding and improving the spatiotemporal dynamics of urban mobility systems for enhanced efficiency, sustainability, and human well-being.

One significant hurdle faced by such mobility initiatives is the difficulty in precisely evaluating the effectiveness and necessity of projects and interventions in alignment with their mobility goals. Such evaluation demands the collection of detailed mobility data across the entire transportation and mobility system. Yet, traditional data collection methods such as traffic surveys, sensor networks, and traffic cameras not only incur significant costs when deployed at scale but also often provide insufficient coverage. These methods typically yield either highly detailed temporal data for narrow segments (e.g., specific areas or intersections) or overly generalized, aggregated data for the whole system. To enable truly effective, data-driven decision-making, there is a critical need for cost-effective methods that can deliver detailed temporal insights spanning the full geographical extent of the mobility system.

\section{Cellular Network Data}
Network data---derived from the cellular network activity of millions of subscribers across extensive areas---can serve as a source of rich mobility information. This data captures the movements and behaviors of large groups, offering insights into the spatial and temporal aspects of population flows within and across regions. The cellular network data explored in this research was obtained from Telia's Crowd Insights platform.

\subsection{Methodology}
To transform cellular signals into mobility data, a network operator collects temporal data generated during routine mobility subscriber activities (such as calls, texts, and movements within network coverage), anonymizing and aggregating this data to respect privacy and meet GDPR standards. Spatial data from network cell coverage is used to estimate geospatial positions, facilitating movement pattern analysis without traditional triangulation. Such a dataset typically excludes sensitive, roaming, and inactive subscriptions, focusing on active mobile subscriptions and extrapolating this information to reflect the broader population in an area. Advanced algorithms distinguish between stationary and moving signals, classifying them as dwells and transits, thus providing a foundation for understanding wider mobility trends via reports such as \cite{Telia2021cim}:

\begin{itemize}
    \item \textbf{Activity Reports}, which shed light on the locations, times, and origins or destinations of groups
    \item \textbf{Trip Reports}, which offer an origin–destination matrix to understand cross-country movement volumes between pairs of locations
    \item \textbf{Routing Reports}, which provide insights into the most likely travel routes taken, encompassing various modes of transportation
\end{itemize}

\subsection{Advantages and Challenges}
As discussed, a serious challenge in analyzing mobility trends is the acquisition of high-resolution temporal data that also spans extensive geographical areas. Cellular network data, collected during routine telecom operations, provides a rich source of mobility information which offers valuable insights into both real-time and historical patterns of population movement. This information is invaluable for researchers, policymakers, and planners in understanding traffic flows, thus facilitating informed, data-driven decisions and policies in urban transportation and mobility planning.

However, leveraging cellular data for mobility analysis comes with some drawbacks and challenges, including variable spatial resolution from differing network coverages and the potential loss of granularity due to anonymization and aggregation. Data processing assumptions, such as stationary signals and proximity to the nearest cell, may not always reflect the complexities of urban mobility. Furthermore, policies excluding data from smaller groups to protect privacy can lead to an underrepresentation of certain areas or times. These limitations necessitate a careful evaluation of the data's utility against the backdrop of privacy considerations and the inherent characteristics of cellular networks.

\section{Routing Reports for Trondheim Municipality}
Routing Reports analyze \textit{peopleFlow}---journey patterns of groups of people---inferring the most probable travel routes to gain insights into travel behaviors, highlighting preferences in routes and transportation modes over various regions and times. This analysis incorporates multiple transportation modes---such as road, rail, ferry, and pedestrian pathways---leveraging OpenStreetMap for mapping probable trip paths over 1km. The routing reports employ an open trip planner to ascertain the quickest route between two points, considering the operator's network coverage and penalizing unnecessary mode switches.

\begin{figure}[t]
    \centering\includegraphics[width=0.55\linewidth]{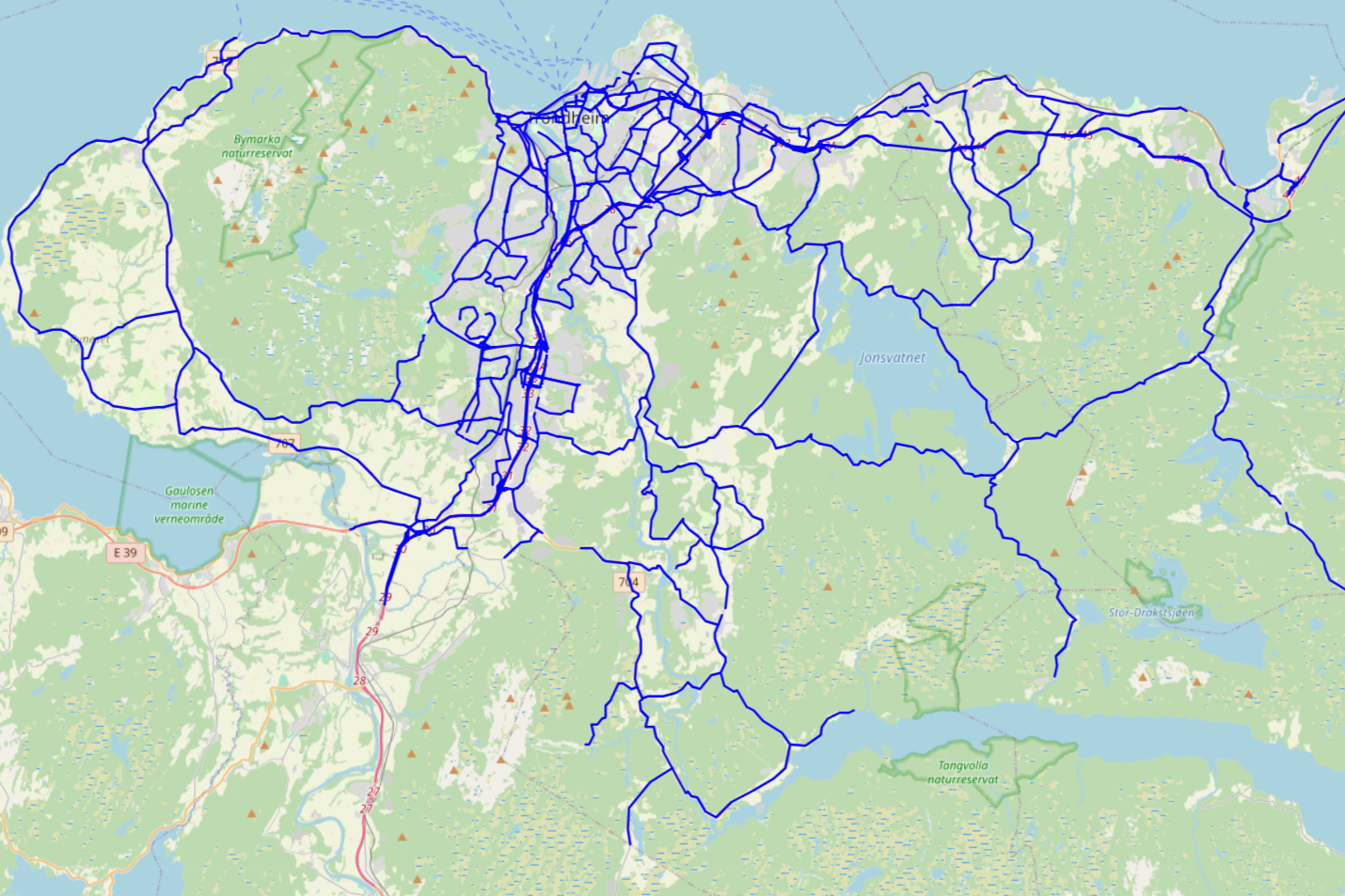}
    \vspace{-10pt}\caption{Satellite map of ways covered by the routing reports}
    \vspace{-10pt}\label{fig:teliaWayIDs}
\end{figure}

\subsection{Data Description}
The routing report dataset used in this study spans from January 17, 2019, to November 30, 2023, covering 1744 dates, with 35 dates omitted due to data gaps or poor signal quality. The raw dataset is aggregated into hourly and daily intervals, resulting in 82,798,311 and 3,792,160 data points respectively. As shown in Fig. \ref{fig:teliaWayIDs}, the dataset geographically covers the entire Trondheim municipality and extends into parts of the neighboring Malvik municipality, due to the location of Trondheim Airport just beyond the municipal boundary. The dataset contains six (or seven) attributes for the daily (or hourly) temporal aggregations, as follows:

\begin{itemize}
    \item \textit{Date and Hour}: These refer to the date and hour the mobility data was recorded. The \textit{Hour} attribute is only present in the hourly temporal aggregation.
    \item \textit{wayID}: This is a unique OpenStreetMap identifier assigned to each road segment. A stretch of road, rail, or ferry route is thus composed of several connected \textit{ways} (segments) each encoded as a \texttt{LineString} object consisting of the longitude and latitude coordinates of its nodes. There are 2210 and 2212 unique \textit{wayID} values in the hourly and daily aggregations, respectively.
    \item \textit{tagKey}: This refers to the transportation mode, one of three options: road, rail, or ferry.
    \item \textit{tagValue}: This is a breakdown of each \textit{tagKey} into its subcategories as defined by OpenStreetMap, such as primary/trunk/secondary for roads and expressboat/cruise for ferry, among others.
    \item \textit{Municipality}: This refers to the municipality the \textit{wayID} falls under based on the geographical coordinates of its nodes, either Trondheim or Malvik.
    \item \textit{peopleFlow}: This is the estimated number of people passing through a specific \textit{wayID} at the daily or hourly aggregation.
\end{itemize}

\subsection{Data Preprocessing, Feature Engineering, and Graph Augmentation}
The raw dataset was preprocessed to rectify missing \ti{tagValue} entries in the hourly and daily aggregations, supplementing them with data from OpenStreetMap. To facilitate direct comparisons, the daily dataset was constrained to \ti{wayIDs} consistent with those in the hourly data. The excluded \ti{wayIDs} from the hourly dataset typically had such low \ti{peopleFlow} volumes in the daily data that the hourly numbers would not meet the minimum threshold of five individuals. In addition, some discrepancies were identified in the daily data, which were corrected to align with the total hourly \ti{peopleFlow} values for the respective days.

Subsequently, the dataset was enriched with temporal attributes by extracting metadata from the \ti{Date} feature---including \ti{Day}, \ti{Month}, \ti{WeekNumber}, \ti{Year}, and \ti{HolidayName}---for subsequent analysis. Historical weather data \cite{VisualCrossing2024} was incorporated to assess the impact of weather conditions on mobility patterns. Additionally, population statistics for Trondheim and Malvik municipalities were sourced from Statistisk Sentralbryå \cite{SSB04861} to examine the relationship between mobility patterns and demographic trends over time.

Routing reports enable the analysis of \ti{peopleFlow} trends by selecting origin and destination coordinates and calculating routes within a mobility network represented as a \texttt{MultiDiGraph}---a graph that allows multiple directed edges between the same pair of nodes. However, the geospatial graph associated with the routing reports has discontinuities, preventing some nodes from being reachable. To address such gaps, a fully-connected graph from OpenStreetMap, covering road, rail, and ferry modes, was obtained and adapted to prioritize ways in the routing reports while ensuring all nodes are accessible. Subsequent trend analyses were then confined to ways (and \ti{wayIDs}) documented in the routing reports and associated \ti{peopleFlow} data.

\fig{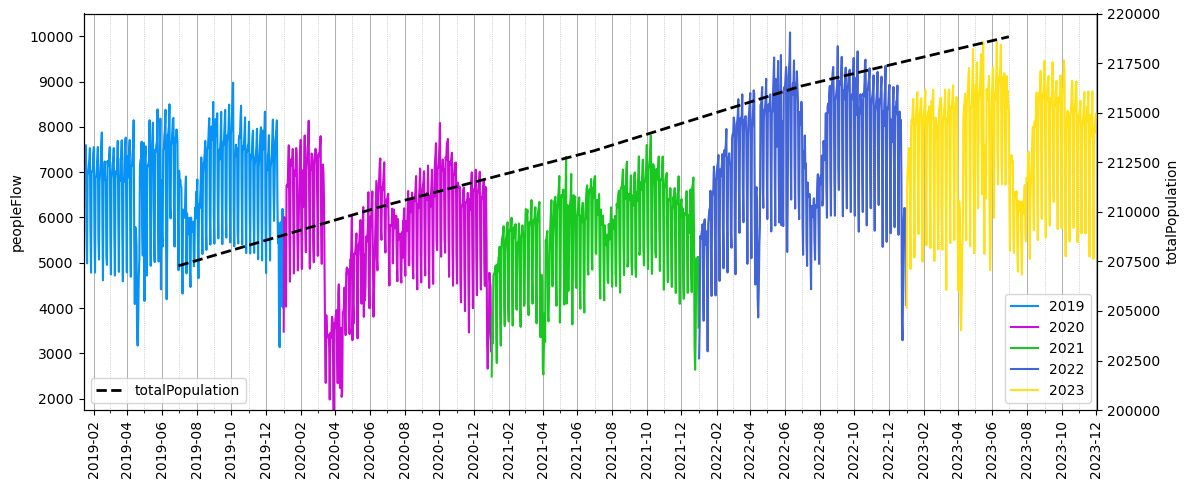}{Yearly breakdown of \ti{peopleFlow} volumes from January 2019 to November 2023.}

\section{Analyses and Discussion}
This section details our investigation into the spatiotemporal dynamics of Trondheim's mobility system, utilizing historical \ti{peopleFlow} volumes obtained from the routing reports. Our analysis encompasses geospatial trends, temporal patterns, and the impact of external factors. In addition, we explore the dynamics of specific routes and areas of interest in the city.

\subsection{Geospatial Trends}
Figure \ref{fig:Date-peopleFlow-totalPopulation-Year} presents the average (mean) \ti{peopleFlow} volumes derived from routing reports between January 7, 2019, and November 30, 2023, juxtaposed with annual population data. This reveals a consistent increase in \ti{peopleFlow} volumes, mirroring the average yearly population growth of approximately 3000 residents. The COVID-19 pandemic's effect is pronounced, showing a dip and subsequent recovery in \ti{peopleFlow} volumes from March 2020 to March 2021. Excluding this period, seasonal declines in \ti{peopleFlow} are evident during Easter, Summer, Christmas, and New Year holidays.
The rationale for using mean \ti{peopleFlow} rather than total sums is due to the overlapping volumes across different routes. For instance, a single individual traversing a road segmented into three ways would be counted in the \ti{peopleFlow} for each segment, thereby overestimating the total mobility if sums were used.

Figure \ref{fig:Date-peopleFlow-Municipality} breaks down the annual \ti{peopleFlow} volumes by municipality, offering insight into the mobility patterns to and from Trondheim. Notably, much of the data from Malvik municipality pertains to road and rail traffic on the E6 trunk highway east of Trondheim which extends to Trondheim Airport. This analysis reveals that Malvik's traffic volumes generally remain lower than Trondheim's, except during significant holiday seasons when there is increased travel in and out of Trondheim.
Furthermore, analyzing the yearly volumes by transportation mode (\ti{tagKey}) in Fig. \ref{fig:Date-peopleFlow-tagKey} highlights periods of increased volume in one mode compared to others. Such fluctuations may signal shifts in the transportation preferences of the city's inhabitants, prompting further investigations.

\fig{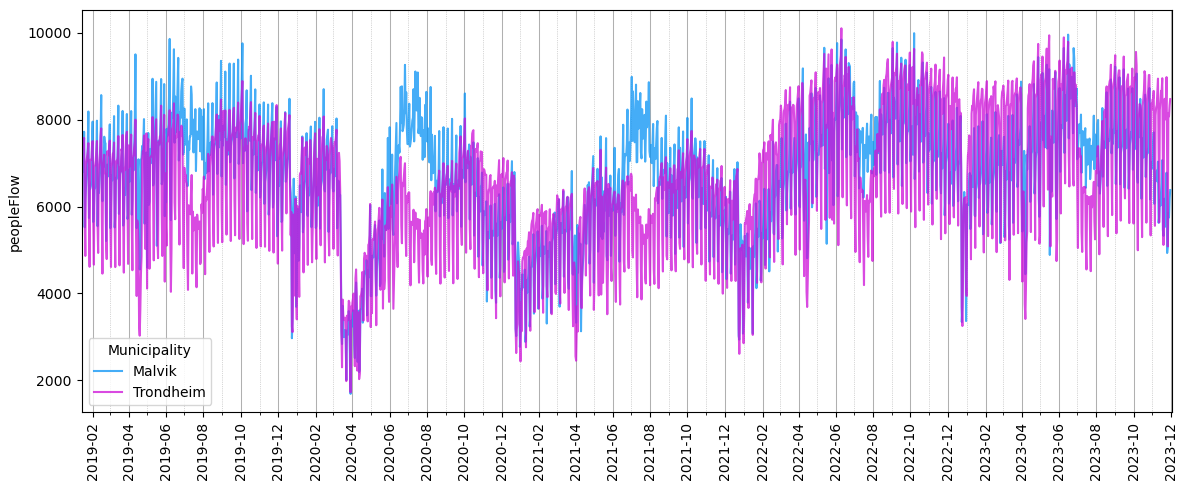}{Detailed overview of \ti{peopleFlow} volumes across municipalities}
\fig{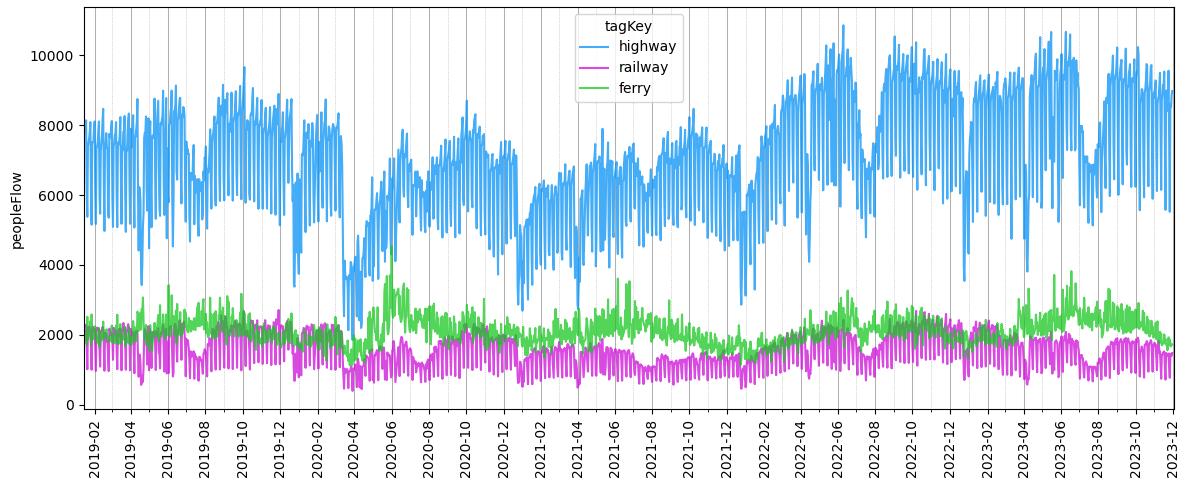}{Detailed overview of \ti{peopleFlow} volumes across transportation modes}

\subsection{Temporal (Normalized) Patterns}
The temporal analysis of the routing reports encompasses hourly, daily, and weekly levels across different transportation modes, focusing on data from March 1, 2022 to November 30, 2023, with the COVID-19 period analyzed separately in Sect. \ref{sec:external-factors}. Due to significant volume disparities among the transportation modes, normalization within the range $[0,1]$ was applied to each mode's data for a fair comparison of flow trends. Figure \ref{fig:Hour-normPeopleFlow-tagKey} shows that the hourly patterns for rail and road traffic are similar, while ferry traffic tends to increase during road and rail's off-peak hours.

Similarly, Fig. \ref{fig:Day-normPeopleFlow-tagKey} reveals that ferry traffic peaks on Sundays, contrasting with the weekend decline in road and rail flows. Meanwhile, Fig. \ref{fig:WeekNumber-normPeopleFlow-tagKey} indicates a less pronounced correlation among the three modes, though road and rail still exhibit parallel trends. Notably, road and rail traffic decrease during the summer holidays, but ferry traffic remains consistent due to tourism around the city. The monthly trends, not shown here, offer a broader view which smoothens out the weekly fluctuations.

\fig{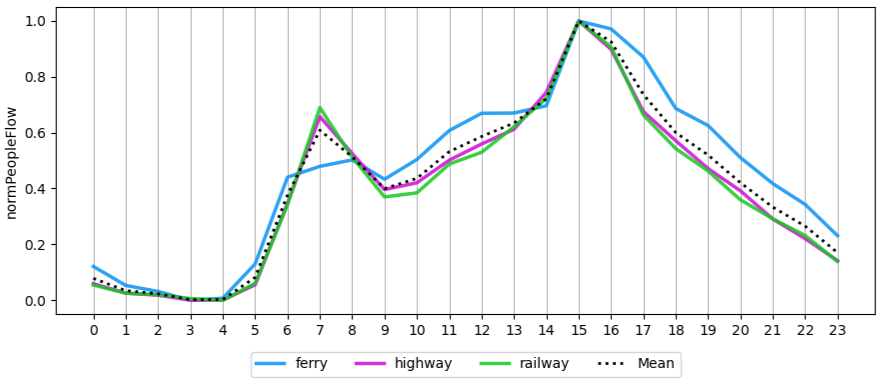}{Normalized hourly variation of \ti{peopleFlow} volumes across transportation modes}
\fig{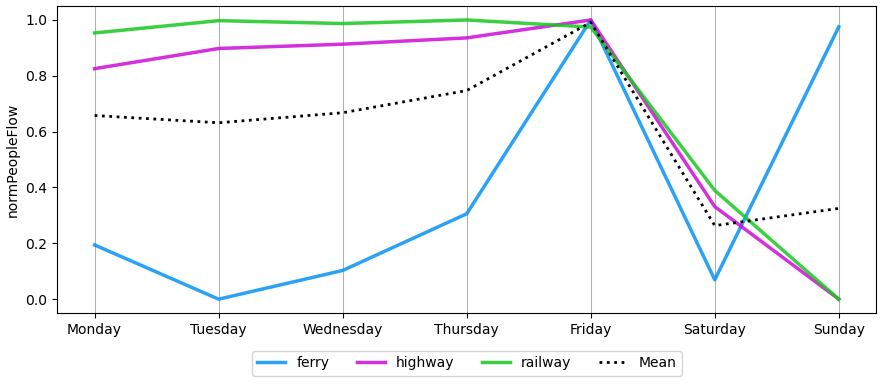}{Normalized daily variation of \ti{peopleFlow} volumes across transportation modes}
\fig{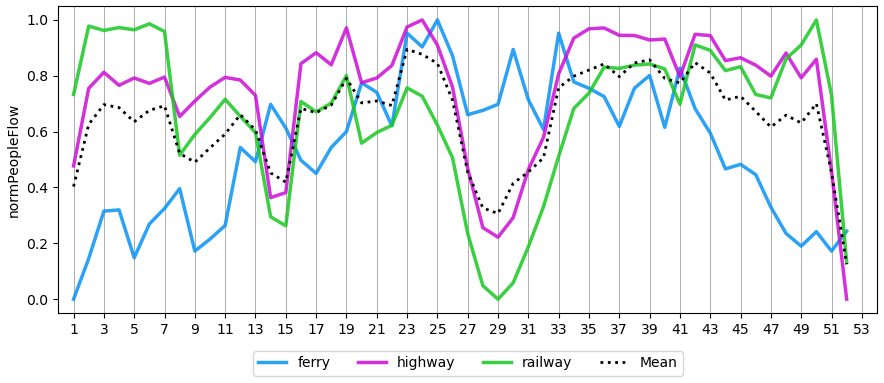}{Normalized weekly variation of \ti{peopleFlow} volumes across transportation modes}

\subsection{External Factors}
\label{sec:external-factors}
This section delves into how external factors like the COVID-19 pandemic, weather conditions, and road attributes (speed limits and lane counts) affect \ti{peopleFlow} volumes across  Trondheim.

\subsubsection{COVID-19 Pandemic}
The pandemic's impact on \ti{peopleFlow} volumes is discernible across Figs. \ref{fig:Date-peopleFlow-totalPopulation-Year}, \ref{fig:Date-peopleFlow-Municipality}, and \ref{fig:Date-peopleFlow-tagKey}. The correlation between the Norwegian government's COVID-19 measures (from March 13, 2020 to March 1, 2022) and \ti{peopleFlow} variations is visible in Fig. \ref{fig:Date-peopleFlow-CovidPeriod}, which presents a 7-day rolling average of the \ti{peopleFlow} volumes. The clear correlation between policy changes and mobility patterns justifies the exclusion of this timeframe from earlier analyses.

\fig{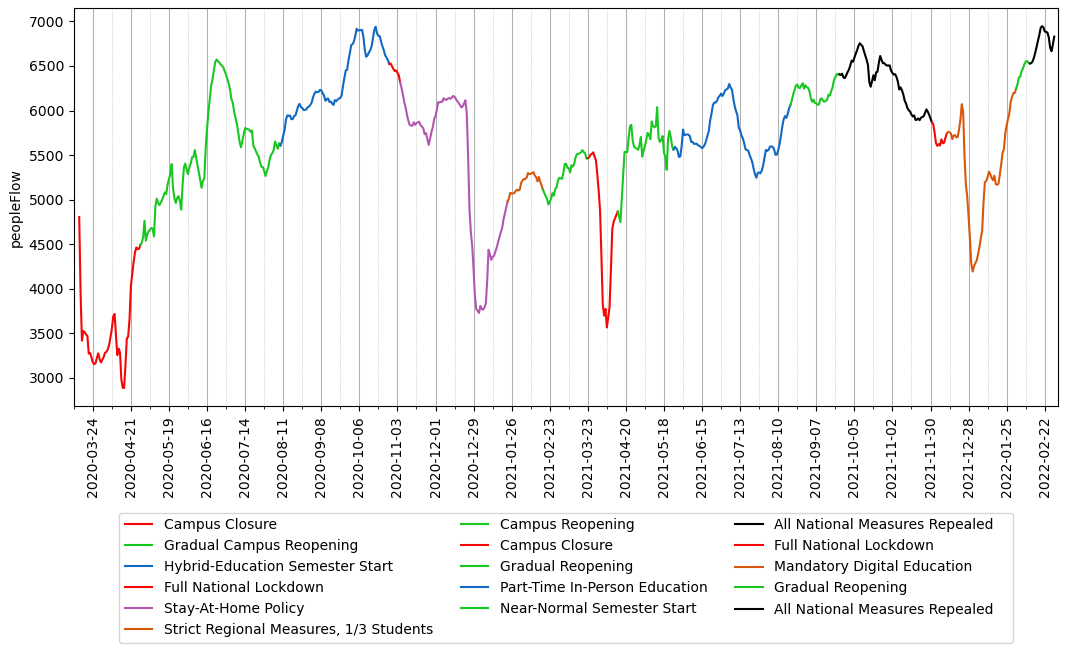}{Effect of COVID-19 measures on \ti{peopleFlow} volumes}

\subsubsection{Weather Conditions}
An analysis incorporating historical weather data shows a slight impact of weather on \ti{peopleFlow} trends. For precipitation types, Fig. \ref{fig:peopleFlow-Precipitation-Conditions-V2} indicates that flow volumes are highest in clear weather, followed by rain, snow, and mixed rain/snow conditions. On the other hand, the lowest flows are associated with rainy, snowy, or cloudy weather conditions.

\fig{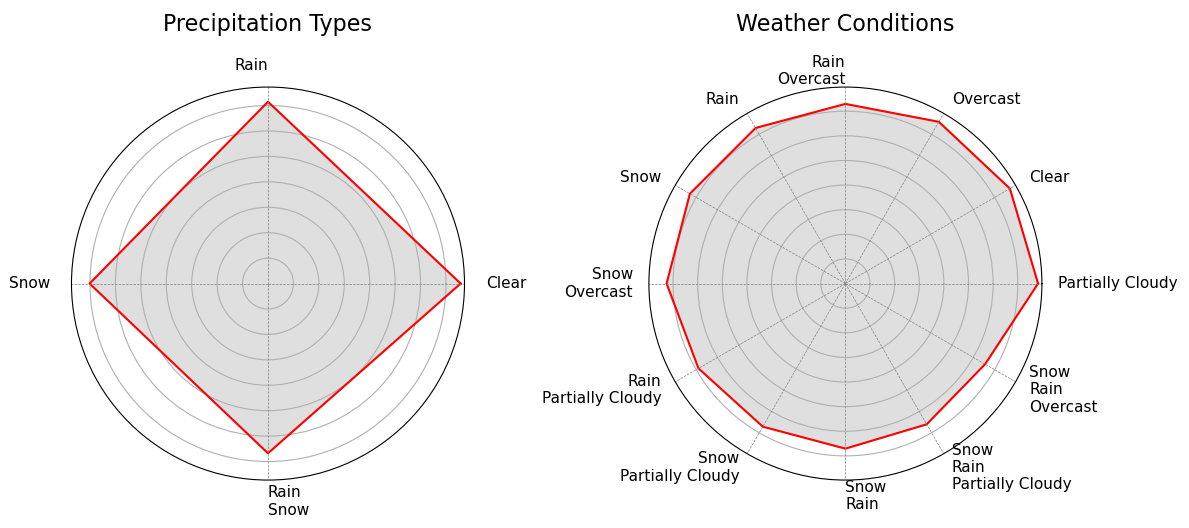}{Variation of \ti{peopleFlow} volumes with weather conditions}

\subsubsection{Speed Limits and Lane Counts}
OpenStreetMap data, linked via \ti{wayIDs} from the routing reports, provided information on speed limits and lane counts for each way. The analysis, depicted in Fig. \ref{fig:peopleFlow-SpeedLimits-Lanes}, reveals a positive correlation between \ti{peopleFlow} volumes and both higher speed limits and greater lane counts. This trend might be influenced by the routing algorithm's preference for quicker routes. Notably, the effect of speed limits on \ti{peopleFlow} plateaus beyond 80km/h due to the reduced number of routes supporting such speeds.

\fig{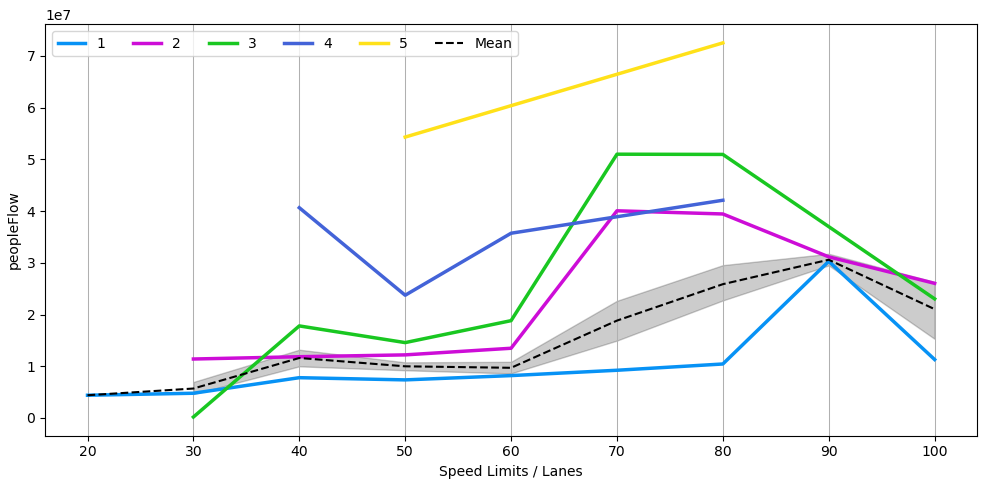}{Combined effect of speed limits and lane count on \ti{peopleFlow} volumes}

\subsection{Specific Routes and Areas}
This section provides a temporal analysis of \textit{peopleFlow} trends along specific routes and areas identified as relevant to the Miljøpakken and MoST initiatives.

\subsubsection{Comparison with Public Transit}
Leveraging Automated Passenger Counting (APC) data from AtB, Trondheim's public transport authority, enabled a comparison between public transit and overall \ti{peopleFlow} volumes along identical routes.

The AtB APC data---covering the period from May 1, 2020 to November 30, 2023---included 1,112,221 unique bus trips over 1295 days, spanning 6 lines and 204 stops. For the analysis in Fig. \ref{fig:peopleFlow-tripSumVolume-Line-atbLines}, both AtB and routing report volumes were smoothed using a 7-day rolling average to eliminate the weekday/weekend variations and subsequently normalized to address the considerable scale differences between them. While a strong correlation exists between AtB and Telia volumes, the instances where AtB volumes surpass Telia volumes stand out and require further investigation, suggesting public transit might be experiencing traffic increases not reflected in the broader mobility system. It is also instructive that the majority of these occurrences took place during the COVID-19 period, analyzed separately earlier.

\fig{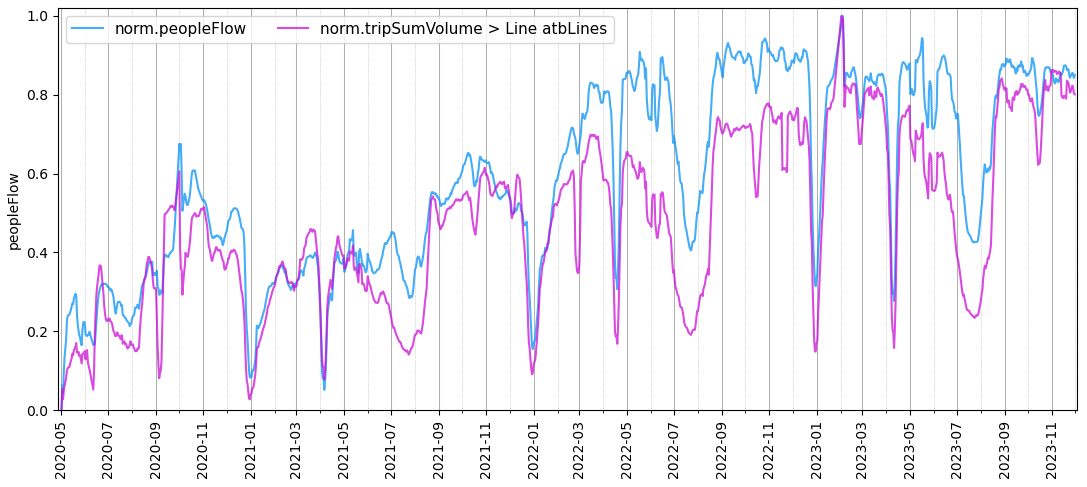}{Comparison of public transit and \ti{peopleFlow} volumes along selected bus routes}

\subsubsection{Miljøpakken Bromstadruta Project}
\begin{figure}[t]
    \centering\includegraphics[width=0.57\linewidth]{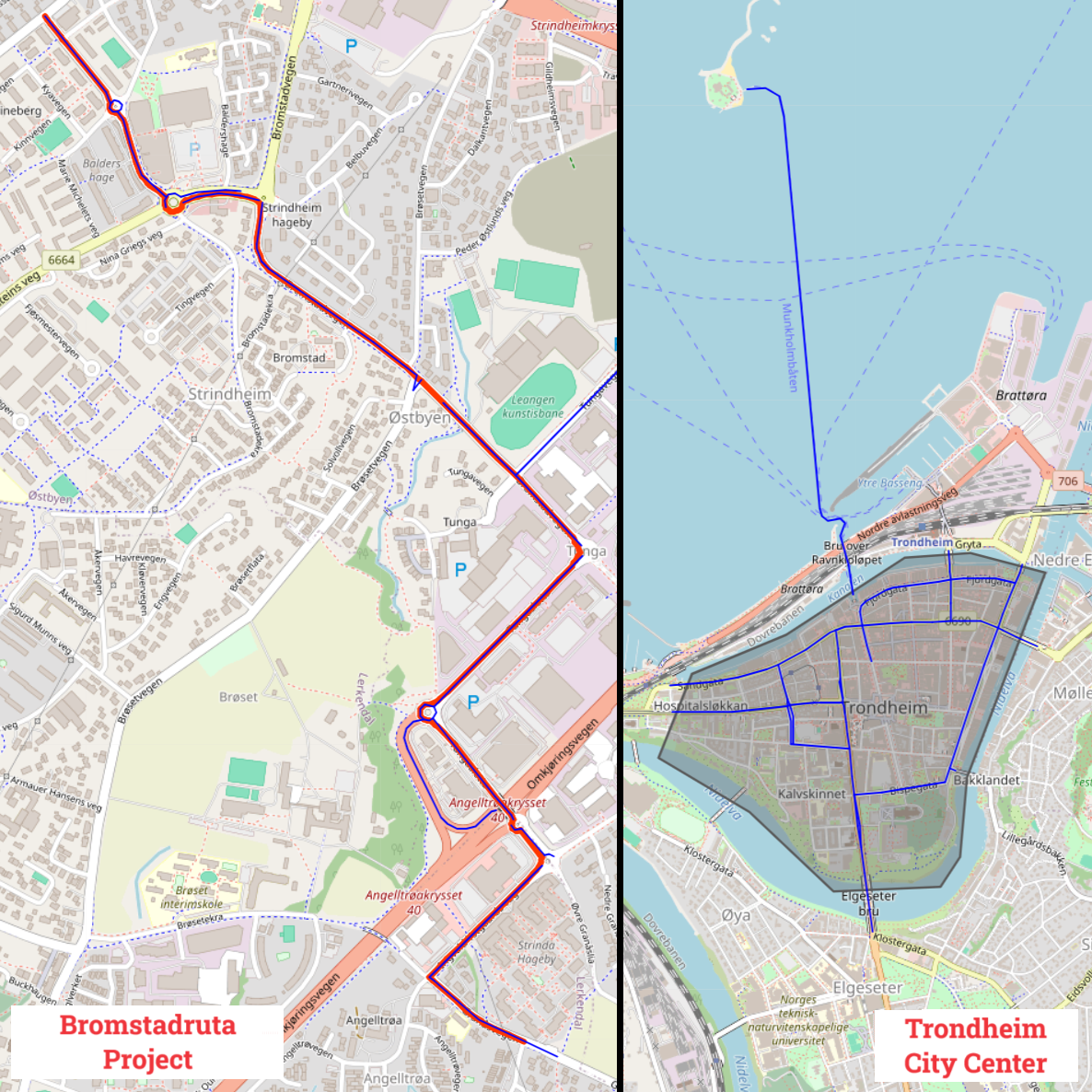}
    \vspace{-10pt}\caption{Bromstadruta cycle path (left) and Trondheim city center (right)}
    \vspace{-10pt}\label{fig:MoST-Bromstadruta-Trondheim-Sentrum-Maps}
\end{figure}

Miljøpakken is developing a 3.2km cycle path and sidewalk project in Trondheim named Bromstadruta \cite{Miljopakken2024bfs}, depicted in Fig. \ref{fig:MoST-Bromstadruta-Trondheim-Sentrum-Maps} (left) with the cycle path outlined in red and the corresponding routing report ways in blue. Such infrastructure projects stand to benefit from a thorough analysis of mobility patterns along the proposed route early in the planning phase. For example, Fig. \ref{fig:peopleFlow-Route-Telia-Bromstadruta-Trondheim} shows normalized \ti{peopleFlow} trends along Bromstadruta versus the city at large, uncovering unique dynamics that could be crucial for planning and achieving project goals.

\fig{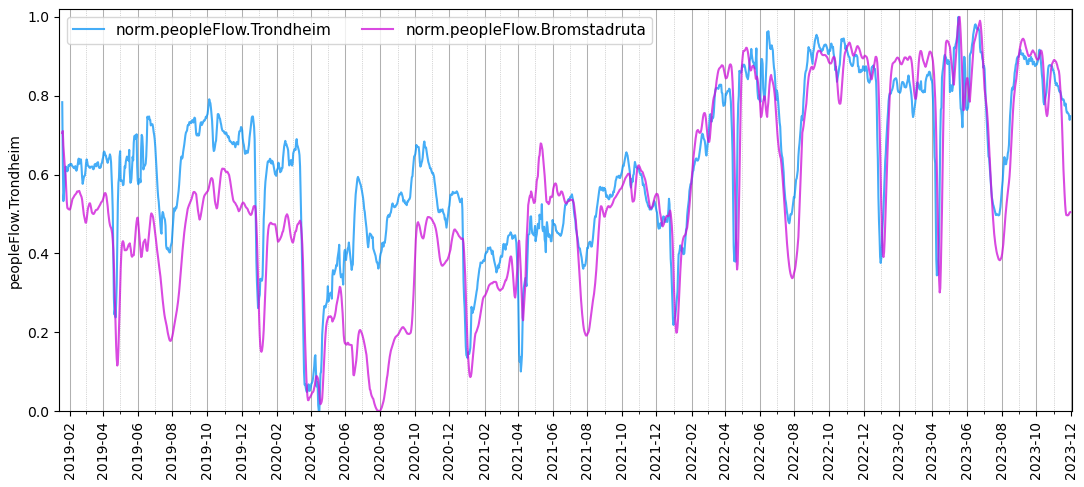}{Analysis of normalized \ti{peopleFlow} trends along the planned Bromstadruta cycle path in Trondheim}

\subsubsection{Trondheim City Center}
A similar analysis for specific urban areas can be conducted, in this case focusing on the city center as shown in Fig. \ref{fig:MoST-Bromstadruta-Trondheim-Sentrum-Maps} (right) with the target area highlighted in gray. Figure \ref{fig:peopleFlow-Route-Sentrum-Island-Ways} reveals the city center's pronounced daily (unnormalized) \ti{peopleFlow} volumes compared to other areas. An in-depth examination of these central urban mobility patterns could provide valuable insights for optimizing traffic flow, managing congestion, and allocating public transit resources more effectively.

\fig{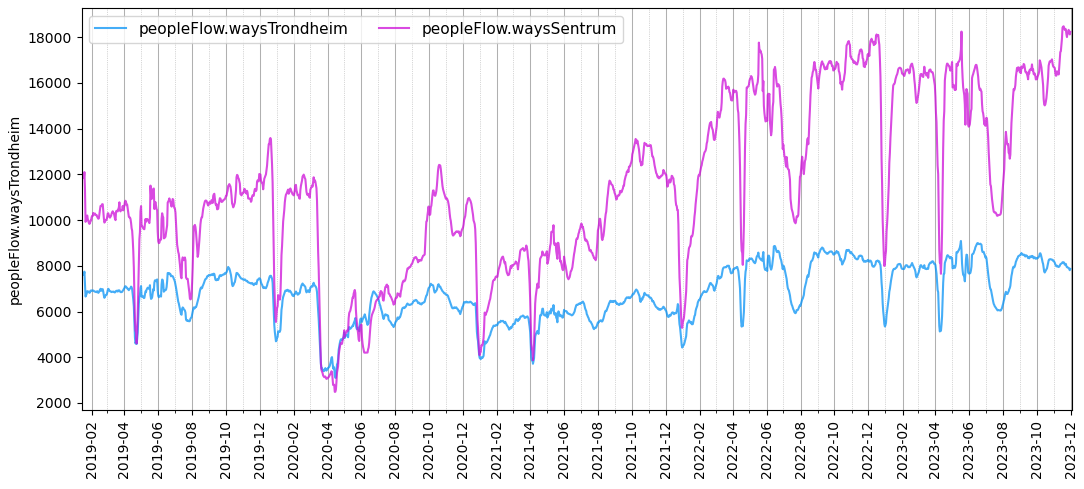}{Analysis of \ti{peopleFlow} trends within Trondheim's city center}

\section{Conclusion and Future Work}
This chapter explores the utility of cellular network data to support efficient and sustainable mobility initiatives like Miljøpakken and MobilitetsLab Stor-Trondheim, offering a cost-effective, detailed, and wide-ranging source of mobility information. Through the analysis of Trondheim, Norway's routing reports, the potential of such data for gaining insights into mobility flows within and across regions becomes evident, highlighting:

\begin{enumerate}
    \item Geospatial trends of the mobility dynamics of specific areas and routes within and across municipalities
    \item Temporal patterns across different transportation modes from across hourly and daily scales
    \item The impact of external factors like pandemics, weather, public transit, and transportation infrastructure on mobility volumes
\end{enumerate}

These insights are crucial for data-driven decision-making in urban mobility planning and policy formulation. However, the inherent nature of cellular network signals and practical realities such as privacy concerns bake some assumptions into the mobility data which require careful consideration in its application. Furthermore, data preprocessing and other enhancements are required to fully capitalize on its potential.

This study is part of a larger initiative to develop a digital twin of Trondheim's transportation infrastructure and mobility system, enriched with comprehensive data-driven historical insights and predictive modelling capabilities. Future research will focus on a thorough spatiotemporal analysis of the mobility network---extending to other modes of transport such as cycling and walking---to pinpoint key ways critical to the \ti{peopleFlow} dynamics across the network.

\subsubsection*{Acknowledgements} This research received funding from the PERSEUS project, a European Union's Horizon 2020 research and innovation program under the Marie Skłodowska-Curie grant agreement No. 101034240. The authors also acknowledge MobilitetsLab Stor-Trondheim (MoST) for their financial contribution and AtB for providing mobility data, which has been instrumental in our research.

\bibliographystyle{splncs04}
\bibliography{bibliography}

\end{document}